\documentclass[aps,prd,reprint,superscriptaddress,nofootinbib,amsmath,amssymb,floatfix]{revtex4-2}

\usepackage[T1]{fontenc}
\usepackage{graphicx}
\usepackage{xcolor}           
\usepackage{booktabs}
\usepackage[colorlinks=true,allcolors=blue]{hyperref}

\newcommand{\mchi}{m_\chi}
\newcommand{\ER}{E_R}

\begin{document}

\title{The neutrino fog for inelastic dark matter}

\author{C. S. Amarasinghe}
\email{chami@ucsb.edu}
\affiliation{Department of Physics, University of California, Santa Barbara, CA 93106, USA}

\begin{abstract}
We characterize the behavior of the neutrino fog for inelastic dark matter scattering on xenon.
The inelastic scattering kinematics reshape the recoil spectrum of the signal and produce a qualitatively different neutrino fog landscape.
In some regions of the inelastic parameter space the fog is completely absent for future xenon experiments at any scale, while in other regions the fog is dispelled only to reappear at larger exposures.
We find that multiple background sources with uncertain normalizations can make the onset of the fog jump discontinuously with the dark matter mass and mass splitting, while the discovery limit varies smoothly.
We also describe a mechanism by which reduced uncertainties on neutrino rates may have the unintuitive effect of raising neutrino floors derived from the fog opacity.
\end{abstract}

\maketitle

\section{Introduction}
\label{sec:introduction}

The first measurements of nuclear recoils from solar $^8$B neutrinos in liquid xenon (LXe)~\cite{XENON:2024ijk, PandaX:2024muv, LZ:2025igz} mark the arrival of the irreducible neutrino background for dark matter searches.
The coherent elastic neutrino-nucleus scattering (CE$\nu$NS)~\cite{Freedman:1973yd} of solar, atmospheric, and supernova neutrinos produces recoil spectra that closely resemble those of Weakly Interacting Massive Particles (WIMPs) of particular masses, and together with uncertainties on the neutrino fluxes this sets a floor on WIMP sensitivity~\cite{Billard:2013qya}.
Low-mass ($\lesssim10$~GeV) WIMP searches have approached the neutrino floor set by solar $^8$B neutrinos.  
The CE$\nu$NS spectra for atmospheric and supernova neutrinos resemble heavier WIMP spectra, and these sources will set the neutrino floor for heavy WIMPs in the next generation of LXe dark matter experiments~\cite{Aalbers:2022dzr, XLZD:2024nsu}.

The floor is not a hard limit, however.
Sensitivity keeps improving with exposure, only more slowly once flux uncertainties rather than Poisson fluctuations dominate the background. 
The neutrino fog framework~\cite{OHare:2021utq}, which we adopt in this work, describes this slowdown without reference to a particular exposure, depending only on the signal and background spectra and their uncertainties.

Neutrino floors and fogs are conventionally mapped for the elastic spin-independent (SI) and spin-dependent (SD) interactions, but SI and SD are not the only viable interactions to consider when designing experiments.
A rich set of models motivate direct detection experiments to search in extended energy regions~\cite{LZ:2026axp,LZ:2023lvz,LZ:2024vge,LUX:2021ksq,XENON:2017fdd,XENON:2022avm,PandaX-II:2017zex,XENON10:2009sho,ZEPLIN-III:2010cnv,CDMS-II:2010wvq,CDEX:2025hez}, beyond the $\sim50$~keV cap suggested by SI/SD scattering. 
Most recently, the LUX-ZEPLIN (LZ) collaboration reported a nuclear recoil candidate at 248~keV in an analysis region where backgrounds are low~\cite{LZ:2026axp}.
Whatever the fate of this candidate, searches in extended energy windows are likely to continue, so characterizing the neutrino fog at extended energies is timely.

We focus on inelastic dark matter that scatters off nuclei into a state that is heavier by a mass splitting $\delta$~\cite{Tucker-Smith:2001myb}. 
Mass splittings of a few hundred keV accommodate the LZ event~\cite{Su:2026rwz,Freese:2026sga,Fan:2026kxx,Rodd:2026tyn,McCabe:2026crm,Fan:2026hzw}. 
The inelastic scattering suppresses the low-energy recoils of the signal and reduces its resemblance to the CE$\nu$NS background, thus reducing the impact of the background as shown for mass splittings of 10 and 50~keV in an energy window up to 50~keV~\cite{Gelmini:2018ogy}. 
We aim for a systematic treatment of the CE$\nu$NS background in an idealized xenon detector searching in an extended energy region for endothermic ($\delta \geq 0$) inelastic dark matter.
Faster dark matter in the halo allows access to larger $\delta$, so in addition to the Standard Halo Model (SHM)~\cite{Baxter:2021pqo}, our analysis considers an additional high-velocity component associated with the Large Magellanic Cloud (LMC)~\cite{Besla:2019xbx}.
Models with momentum-dependent interactions~\cite{Chang:2009yt, Feldstein:2009tr, Fitzpatrick:2012ix} can also produce events at high recoil energies, but we do not consider them here. 

We find that the fog clears entirely above $\delta \approx 180$~keV in both halo models, well inside the kinematic limit for most masses, so searches for splittings of a few hundred keV are limited not by the neutrino background but by exposure and reducible backgrounds. 
At smaller splittings the fog has a separate feature for each neutrino source. 
We describe how floors derived from the fog can jump as the signal shape changes, and can even rise when the uncertainty on one neutrino source is reduced.

Section~\ref{sec:spectra} describes the signal and background spectra and Sec.~\ref{sec:stats} the
statistical method.
Section~\ref{sec:results} maps the fog for inelastic dark matter, Sec.~\ref{sec:floors} describes two effects on fog-derived neutrino floors, and Sec.~\ref{sec:discussion} discusses the implications.
We conclude in Sec.~\ref{sec:conclusion}.

\section{Recoil energy spectra}
\label{sec:spectra}

\subsection{Inelastic dark matter}

We compute dark matter spectra with the WimPyDD package~\cite{Jeong:2021bpl}.
The differential rate is
\begin{equation}
\begin{split}
\frac{dR}{d\ER} = {} & \frac{\rho_0\, \sigma_0}{2\mchi \mu^2}\, F^2(\ER) \\
& \times \int_{v > v_{\min}(\ER)} \frac{f(\vec{v})}{v}\, d^3v,
\end{split}
\label{eq:dm_rate}
\end{equation}
where $\rho_0=0.3$~GeV/cm$^3$ is the local dark matter density, $\mchi$ and $m_N$ are the dark matter and target nucleus masses, $\mu$ is their reduced mass, $\sigma_0$ is a reference dark-matter-nucleon cross section scaled to the nucleus, $F(\ER)$ is the nuclear form factor, and $f(\vec{v})$ is the dark matter velocity distribution in the detector frame.
We use the default WimPyDD nuclear response function, based on the shell-model calculation of Ref.~\cite{Anand:2013yka}.
The lower bound $v_{\min}(\ER)$ is the minimum speed required to produce a recoil of energy $\ER$.
For elastic scattering $v_{\min} = \sqrt{m_N \ER/ (2\mu^2)}$, with $\mu = \mchi m_N/(\mchi + m_N)$.

Dark matter that scatters inelastically into a heavier state arises for nearly pure higgsinos and other models~\cite{Tucker-Smith:2001myb,Nagata:2014wma}.
The minimum speed becomes
\begin{equation}
   v_{\rm min}(\ER,\delta) = \frac{1}{\sqrt{2 m_N \ER}}\left(\frac{m_N \ER}{\mu} + \delta\right),
   \label{eq:vmin_idm}
\end{equation}
where $\delta$ is the mass splitting between the outgoing and incoming states.
This has a minimum of $\sqrt{2\delta/\mu}$ at $\ER = \mu\delta/m_N$, so the spectrum is peaked at non-zero recoil energy and vanishes below a kinematic threshold.
For a given interaction strength the integrated rate falls steeply with $\delta$, following the tail of the velocity distribution, and scattering is forbidden altogether above values of $\delta$ that require velocities that are too high to be available in the halo.

\begin{figure}
    \centering
    \includegraphics[width=\columnwidth]{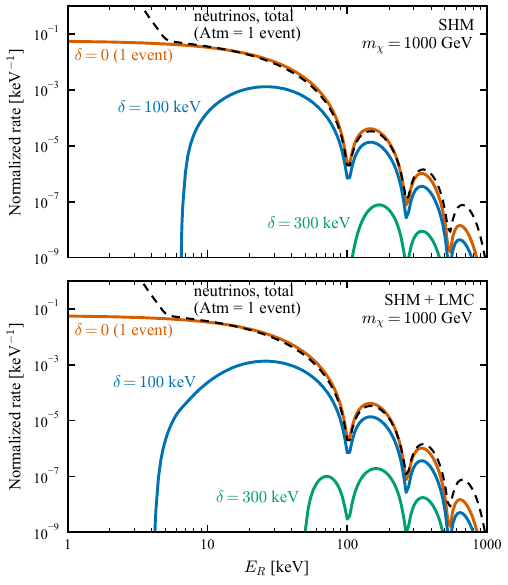}
    \caption{Recoil spectra at $m_\chi = 1000\,\mathrm{GeV}$ for the SHM (top) and SHM + LMC (bottom). The
  $\delta = 0$ signal is normalized to one event and the $\delta > 0$ signals keep their relative
  rates. The dashed line is the total neutrino background, normalized so that the atmospheric
  component is one event.}
    \label{fig:recoils}
\end{figure}

The signal at large $\delta$ is therefore supplied entirely by the high-speed tail of $f(\vec{v})$.
We use two benchmark halo models, and the iDM recoil energy spectra under both models are shown in Fig.~\ref{fig:recoils}.
The first is the Standard Halo Model (SHM) described in Ref.~\cite{Baxter:2021pqo}, the recommended convention for reports in direct detection.
The second adds unvirialized high-speed components from dark matter associated with the Large Magellanic Cloud (LMC)~\cite{Rodd:2026tyn,Smith-Orlik:2023kyl,Besla:2019xbx} after its recent merger with the Milky Way.
Following Ref.~\cite{Rodd:2026tyn}, both SHM and LMC are modeled as Gaussian velocity distributions in the Galactic frame, truncated at a maximum speed,
\begin{equation}
\begin{split}
f(\vec v) = {} & (1-w)\,\frac{e^{-|\vec v|^{2}/v_0^{2}}}{N_{\rm SHM}}\,
    \Theta\!\left(v_{\rm esc}-|\vec v|\right) \\
  & + w\,\frac{e^{-|\vec v-\vec v_b|^{2}/\sigma_b^{2}}}{N_{\rm LMC}}\,
    \Theta\!\left(v_{\rm cut}-|\vec v-\vec v_b|\right),
\end{split}
\label{eq:halo}
\end{equation}
where each term is normalized to unit density and $w$ is the fraction of the local dark matter density carried by the LMC component.
For our results we take $w=0$ (pure SHM) and $w=0.6\%$ (SHM+LMC) as benchmarks~\cite{Rodd:2026tyn}.
The SHM is centered on the Galactic rest frame with dispersion $v_0$ and truncated at the Galactic escape velocity $v_{\rm esc}$.
The LMC component is a Gaussian centered on a bulk velocity $\vec v_b$ at an angle $\cos\beta$ with the Sun's velocity, with dispersion $\sigma_b$ and truncation at $|\vec v-\vec v_b|=v_{\rm cut}$.
Table~\ref{tab:halo} gives the numerical values used in our simulation.
Both distributions are boosted to the laboratory frame by the Sun's Galactic velocity $\vec v_\odot$. We do not include the Earth's orbital motion or any time dependence.

The LMC component increases the reach of Earth-based experiments for iDM.
As the kinematic limit scales as $\delta_{\max}=\mu\,v_{\max}^{2}/2$, the LMC component raises it by a factor of $1.5$.
For $m_\chi=1$~TeV scattering on $^{136}$Xe, the maximum accessible $\delta$ for the SHM and LMC is $395$~keV and $586$~keV, respectively.
Below $\delta\sim100$~keV the two halos give similar spectra to within a few percent.
Above it the LMC component supplies an increasing share of the rate, and beyond the SHM's limit, all of it.

\begin{table*}[t]
  \centering
  \caption{
  Parameters of the halo models.
  Velocities are in the Galactic frame with components ordered (toward the Galactic center, direction of rotation, Galactic north). 
  The LMC component is added to the SHM with weight $w$. 
  $\vec v_\odot$ is the sum of the
  local circular velocity $(0,\,238,\,0)$ and the Sun's peculiar velocity $(11.1,\,12.2,\,7.3)$, which are the only velocities included; Earth's orbital motion is not considered. 
  The kinematic limit $\delta_{\max}=\mu v_{\max}^{2}/2$ is for $^{136}$Xe.}
  \label{tab:halo}
  \begin{tabular}{@{}l c c@{}}
    \toprule
                                                     & SHM                        & LMC component \\
    \midrule
    Fraction of local density                        & $1-w$                      & $w = 0.6\%$ \\
    Center $\vec v_c$ [km/s]                         & $0$                        & $|\vec v_b| = 570$ \\
    Width $v_0,\ \sigma_b$ [km/s]                    & $238$                      & $100$ \\
    Truncation [km/s]                                & $|\vec v|<v_{\rm esc}=544$ & $|\vec v-\vec v_b|<v_{\rm cut}=200$ \\
    Angle to $\vec v_\odot$, $\cos\beta$             & ---                        & $-0.71$ \\
    \midrule
    Sun's velocity $\vec v_\odot$ [km/s]             & \multicolumn{2}{c}{$(11.1,\ 250.2,\ 7.3)$} \\
    $|\vec v_\odot|$ [km/s]                          & \multicolumn{2}{c}{$250.55$} \\
    \midrule
    Sun's speed in the component frame [km/s]        & $250.55$                   & $768.4$ \\
    Largest laboratory speed $v_{\max}$ [km/s]       & $794.6$                    & $968.4$ \\
    $\delta_{\max}$, $m_\chi = 1$~TeV, $^{136}$Xe [keV]      & $395$                      & $586$ \\
    \bottomrule
  \end{tabular}
\end{table*}

\subsection{Neutrinos}
\label{sec:nu_sources}

\begin{figure}
    \centering
    \includegraphics[width=\columnwidth]{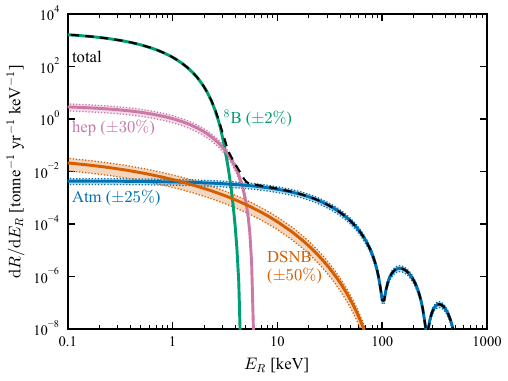}
    \caption{Neutrino backgrounds in a xenon target as a function of nuclear recoil energy: $^8$B, \textit{hep}, atmospheric (Atm) and the diffuse supernova background (DSNB), with their sum (dashed).
    No efficiency or threshold is applied.
    Each shaded band, bounded by dotted lines, is the $1\sigma$ normalization uncertainty $\xi_j$ of Table~\ref{tab:fluxes}; the 2\% uncertainty on the $^8$B spectrum is smaller than the line thickness.
    }
    \label{fig:neutrino_background}
\end{figure}

We consider $^8$B and \textit{hep} solar neutrinos, the diffuse supernova neutrino background (DSNB), and atmospheric neutrinos.
The differential recoil rate from each source is
\begin{equation}
\begin{split}
\frac{dR}{d\ER} = {} & N_T \int_{E_\nu^{\min}(\ER)}^{E_\nu^{\max}}
\frac{d\Phi_\nu}{dE_\nu} \\
& \times \frac{d\sigma}{d\ER}(E_\nu, \ER)\, dE_\nu,
\end{split}
\label{eq:cevns_rate}
\end{equation}
where $N_T$ is the number of target nuclei, $d\Phi_\nu/dE_\nu$ is the neutrino flux spectrum, and $E_\nu^{\min} = \sqrt{m_N \ER/2}$ is the minimum neutrino energy that can produce a recoil of energy $\ER$.
We use the flux spectra tabulated in the public repository of Ref.~\cite{OHare:2021utq}.
The solar fluxes follow the Barcelona 2016 GS98 Standard Solar Model~\cite{Vinyoles:2016djt}, with the $^8$B normalization and its uncertainty taken from a global fit to solar neutrino data~\cite{Bergstrom:2016cbh}.
The DSNB follows Ref.~\cite{Beacom:2010kk}, and the atmospheric flux is the FLUKA calculation of Ref.~\cite{Battistoni:2005pd}.
The cross section is
\begin{equation}
\frac{d\sigma}{d\ER} = \frac{G_F^2 m_N Q_W^2}{4\pi}
\left(1 - \frac{m_N \ER}{2E_\nu^2}\right) F^2(\ER),
\label{eq:cevns_xsec}
\end{equation}
where $G_F$ is the Fermi constant and $Q_W = N - \left(1 - 4\sin^2\theta_W\right) Z$ is the weak nuclear charge for $N$ neutrons and $Z$ protons.
We adopt the same shell-model form factor as Eq.~\eqref{eq:dm_rate}~\cite{Anand:2013yka}, which weights protons and neutrons equally.
Since CE$\nu$NS couples predominantly to neutrons, this neglects differences between the neutron and proton distributions~\cite{Hoferichter:2020osn}.
Using the weak form factor for CE$\nu$NS instead lowers our fog-derived floors by up to a factor of 4 at representative points but leaves the fog-free edge unchanged, so our choice is conservative.
The calculation as a whole is nonetheless idealized, assuming perfect energy resolution and unit efficiency over the full recoil range favor separating signal from background, so the fog in a real detector may be larger.
The flux normalisations and their fractional uncertainties $\xi_j$ listed in Table~\ref{tab:fluxes} are obtained from Ref.~\cite{OHare:2021utq} and the references therein.
Recoil energy spectra for each source and the associated rate uncertainties are shown in Fig.~\ref{fig:neutrino_background}.

\begin{table}[tbp]
    \centering
    \caption{Neutrino sources, flux normalisations and the fractional uncertainties $\xi_j$ used in
    this work. $\xi_j$ is the width of the Gaussian prior on the normalisation of source $j$ in the
    profile likelihood. The DSNB flux is the all-flavor flux.}
    \label{tab:fluxes}
    \begin{tabular}{@{}lcc@{}}
      \toprule
      Source & Flux [cm$^{-2}$\,s$^{-1}$] & $\xi_j$ \\
      \midrule
      $^{8}$B       & $5.16\times10^{6}$ & $0.02$ \\
      \textit{hep}  & $7.98\times10^{3}$ & $0.30$ \\
      DSNB          & $85.7$             & $0.50$ \\
      Atmospheric   & $10.7$             & $0.25$ \\
      \bottomrule
    \end{tabular}
\end{table}

\section{Statistics}
\label{sec:stats}

We follow the neutrino fog calculation introduced in Ref.~\cite{OHare:2021utq}.
The discovery limit $\sigma$ is the smallest cross section for which the background-only hypothesis is rejected at $3\sigma$ in the median experiment.
It is obtained from a profile likelihood ratio test using a binned Poisson likelihood in recoil energy, multiplied by a Gaussian constraint on the normalization of each neutrino flux with fractional width $\xi_j$.
The significance is $Z=\sqrt{q_0}$ from the asymptotic distribution of the test statistic~\cite{Cowan:2010js}.
We bin the recoil energy spectra between $10^{-4}$\,keV and $10^{3}$\,keV in 400 logarithmically spaced bins.
The efficiency is unity in this energy range, and we assume perfect energy resolution.
The four background normalizations ($^8$B, \textit{hep}, atmospheric and DSNB) are the only nuisance parameters, each an independent Gaussian with width $\xi_j$ and no shape uncertainty.
The dark matter mass $m_\chi$ and mass splitting $\delta$ are fixed in each test.

The scaling of the discovery limit with exposure follows from the number of background counts $N$~\cite{Billard:2013qya,OHare:2021utq}.
For $N \ll 1$ the search is background-free and $\sigma \propto 1/N$.
With a background whose spectrum resembles the signal, the limit improves by Poissonian subtraction, $\sigma \propto 1/\sqrt{N}$.
A systematic rate uncertainty on the background stalls this improvement as $\sigma \propto \sqrt{({1 + \xi^2 N})/N}$, which saturates for $N \gtrsim 1/\xi^2$.
The limit resumes improving only when the exposure is large enough to resolve the difference in spectral shape between signal and background.

The opacity of the fog is the index of this scaling~\cite{OHare:2021utq},
\begin{equation}
    n = - \left(\frac{d \ln \sigma}{d \ln N} \right)^{-1},
    \label{eq:opacity}
\end{equation}
so that $n=1$ in the background-free regime, $n=2$ under Poissonian subtraction, and $n>2$ when the limit is stalled by normalization uncertainty.
The fog is the region with $n>2$.
We denote by $N_\star$ the number of background counts at which $n=2$ is first exceeded, by $\sigma_\star$ the corresponding discovery limit cross section, and by $n_{\max}$ the largest opacity reached at any exposure.
A model with $n_{\max} \le 2$ never enters the fog.

\section{Neutrino fog for inelastic dark matter}
\label{sec:results}

\begin{figure}
    \centering
    \includegraphics[width=\columnwidth]{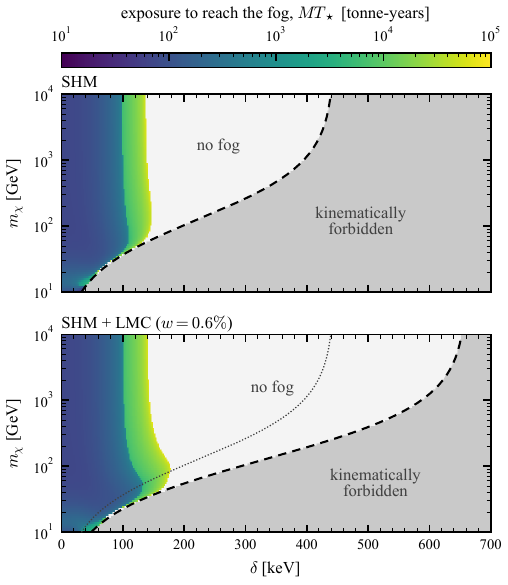}
    \caption{Exposure needed to reach the floor derived from the xenon neutrino fog ($n=2$), $MT_\star$ in tonne-years for the SHM (top) and the SHM with the LMC component ($w=0.6\%$, bottom).
    In the white regions $n$ never exceeds 2 at any exposure, and the gray regions are kinematically forbidden (the SHM limit is repeated as a dotted line in the bottom panel).
    The abrupt change in color near $\delta\simeq100$\,keV corresponds to around a factor 10 change in required exposure.
    }
    \label{fig:heatmap}
\end{figure}

The mass splitting $\delta$ moves the signal to higher recoil energies (see Fig.~\ref{fig:recoils}), above most of the atmospheric spectrum, and further into a range where no neutrino backgrounds are degenerate with the signal.
Figure~\ref{fig:heatmap} shows the exposure required to reach the fog for inelastic dark matter scattering into heavier states ($\delta \geq 0$) in SHM and SHM+LMC scenarios.
For elastic scattering ($\delta=0$) the fog arrives at about 70~tonne-years for $m_\chi \gtrsim 100$~GeV, and the required exposure grows smoothly with $\delta$ except for an abrupt step.
The step corresponds to where the fog onset moves from an opacity peak set by the atmospheric component to an opacity peak set by a combination of the atmospheric and DSNB components, and creates an order of magnitude change in required exposure near $\delta \simeq 100$~keV.
The opacity peaks will be described below.

At $m_\chi = 1$~TeV the fog is completely absent beyond $\delta \approx 140$~keV up to the highest exposures we test ($10^{10}$~tonne-years), and at any mass beyond 146~keV (178~keV) for the SHM (LMC).
A wide region of the parameter space that is still within the kinematic limit therefore has no neutrino fog.
The LMC component extends both the kinematic reach and the region with fog to larger $\delta$, without changing the picture.
For masses below around 50~GeV, the fog persists almost to the kinematic limit, so regions of no fog only arise for heavier inelastic dark matter.

\begin{figure*}
    \centering
    \includegraphics[width=\textwidth]{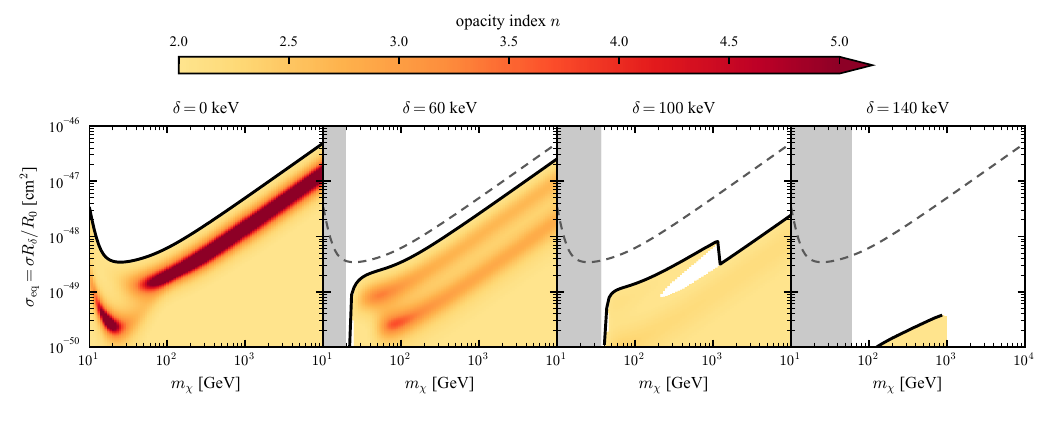}
    \caption{The neutrino fog for four mass splittings and the SHM.
    White above the black line has $n<2$, where the discovery limit improves unimpeded with exposure.
    The solid black line is the fog onset $\sigma_\star$, the cross section at which $n$ first crosses 2, and the dashed grey line repeats $\sigma_\star$ at $\delta=0$ in every panel.
    Grey bands are masses with no signal at that $\delta$ (kinematically forbidden).
    The vertical axis is the cross section rescaled to produce the same number of signal events as in the elastic ($\delta=0$) case.
    The step of $\sigma_\star$ near $m_\chi\simeq1$\,TeV at $\delta=100$\,keV is the first peak of $n(N)$ falling below $n=2$, and the white sliver below the line at lower masses is the valley between the two peaks.
    At $\delta=140$\,keV the fog is found only for $m_\chi\lesssim1$\,TeV.}
    \label{fig:landscape}
\end{figure*}

Figure~\ref{fig:landscape} shows the xenon neutrino fog for inelastic dark matter (SHM) in the familiar $(\mchi, \sigma)$ plane, with the cross sections rescaled to give equal signal counts for varying $\delta$.
The cross section is scaled by the inelastic to elastic rate ratio $R_\delta/R_0$ (which is less than unity due to the loss of support from the halo velocity distribution), to isolate the effect of differences in signal shape from the effect of signal rate.
The shape differences lead to a lowering of the neutrino floor derived from the fog -- the cross section at which $n=2$ is first crossed.
The opacity develops a two-peak structure that leads to a jump in the neutrino floor around ($m_\chi=1$~TeV, $\delta=100$~keV) as the first peak dips below the $n=2$ threshold.
By $\delta=140$~keV, the fog is absent above around 1~TeV.

\begin{figure}
    \centering
    \includegraphics[width=\columnwidth]{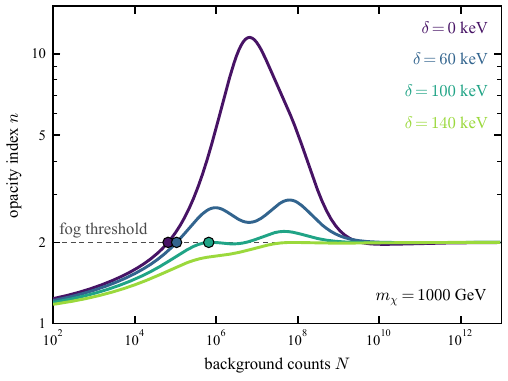}
    \caption{Opacity index $n$ as a function of the number of background counts $N$, for $m_\chi = 1000$\,GeV, the SHM, and four mass splittings.
    The dots mark the first crossing into the fog $N_\star$.
    The fog becomes more transparent as $\delta$ grows owing to the increasing recoil shape differences.
    At $\delta = 140$\,keV, $n$ is always less than 2 up to the largest exposure scanned ($10^{10}$ tonne-years) and no neutrino floor exists.}
    \label{fig:nvsN}
\end{figure}

The opacity for $m_\chi=1$~TeV and several values of $\delta$ is shown in Fig.~\ref{fig:nvsN}.
For elastic scattering the signal spectrum closely resembles the atmospheric spectrum, so the atmospheric normalization uncertainty can absorb the signal over a wide range of exposures, producing a single tall peak in $n(N)$.
As $\delta$ increases, the kinematic threshold removes the low-energy part of the signal, where most of the atmospheric spectrum lies, so this peak shrinks rapidly.
A second peak emerges at higher exposures. 
Raising the atmospheric normalization to absorb the signal over-predicts counts at low recoil energies, which the fit offsets by lowering the DSNB normalization.
The combination resembles the signal better than the atmospheric spectrum alone.
This peak therefore shrinks more slowly with $\delta$, and near $\delta \simeq 100$~keV it takes over from the first peak in setting the fog onset.
At $\delta=100$~keV, the valley between the peaks dips below $n=2$, producing the sliver of transparency in Fig.~\ref{fig:landscape}.
At $\delta=140$~keV, neither opacity peak reaches $n=2$, and the region is completely fog-free.

\section{Limitations of opacity-based neutrino floors}
\label{sec:floors}

A neutrino floor defined as the onset of the fog (the first crossing of $n = 2$) inherits from the continuous but multi-peaked function $n(N)$, and this has two consequences that do not arise for elastic scattering.
We describe these effects so that future work employing the neutrino fog calculation can check that floors derived from the fog behave as expected from a standard fixed-exposure discovery limit calculation.

\subsection{The fog-derived floor jumps with $m_\chi$ and $\delta$}

\begin{figure}
    \centering
    \includegraphics[width=\columnwidth]{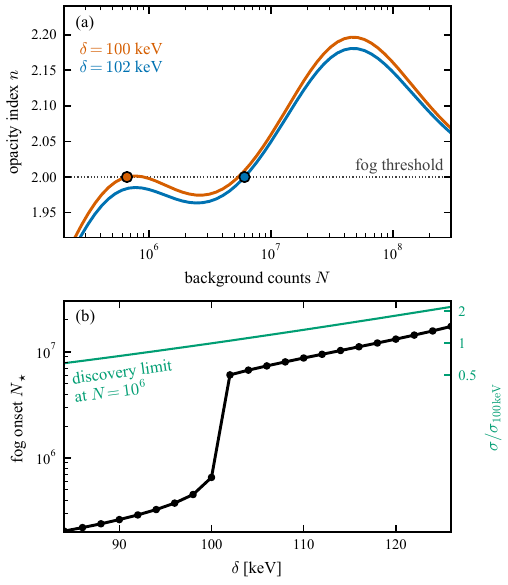}
    \caption{Jump of the fog onset at $m_\chi = 1000$\,GeV (SHM) as $\delta$ increases. (a)~There are two peaks in the opacity index $n$ against background counts $N$ for $\delta = 100$ and $102$\,keV.
    As $\delta$ increases, the first crossing of $n = 2$ (dots), which defines the onset $N_\star$, moves from the rising edge of the first peak to that of the second.
    (b)~The fog onset $N_\star$ against $\delta$ (black, left axis) rises by an order of magnitude in a $\sim$2~keV step of $\delta$.
    The smooth discovery limit at fixed $N = 10^{6}$, between the two onsets, is shown normalised to its value at $\delta = 100$\,keV (green, right axis).
    }
    \label{fig:jump}
\end{figure}

When two or more neutrino sources contribute to the signal region, $n(N)$ has multiple peaks, and the onset is set by whichever peak first crosses $n = 2$.
We encountered a jump in the fog-derived floor in the $\delta=100$~keV panel of Fig.~\ref{fig:landscape}, caused by $m_\chi$ lifting the shape degeneracy and leading the first peak to suddenly miss the $n=2$ threshold.
Jumps arising from the shape-degeneracy lifting are also caused by $\delta$, and in general any shape-varying parameter.
Figure~\ref{fig:jump} shows an order of magnitude jump in the required exposure to reach the neutrino floor over a 2~keV step in $\delta$.
The discovery limit calculated for a fixed exposure changes smoothly over $\delta$, so the discontinuity is a feature arising from the threshold definition and not from the statistical analysis.

\subsection{Lower neutrino uncertainties can raise the fog-derived floor}

\begin{figure}
    \centering
    \includegraphics[width=\columnwidth]{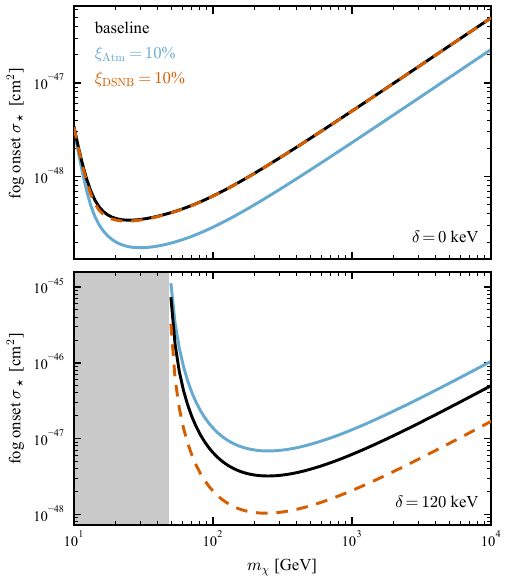}
    \caption{The cross section at the fog onset, $\sigma_\star$, against dark matter mass for the SHM, with the neutrino rate uncertainties at their baseline values $\xi_{\rm Atm} = 25\%$ and $\xi_{\rm DSNB} = 50\%$ (black), with the atmospheric uncertainty reduced to 10\% (blue), and with the DSNB uncertainty reduced to 10\% (orange), all other uncertainties unchanged.
    Top: elastic scattering ($\delta = 0$), where the reduced atmospheric uncertainty lowers the floor.
    Bottom: $\delta = 120$\,keV, where the atmospheric uncertainty reduction raises the floor at every mass by a factor of up to 2, and the DSNB uncertainty reduction lowers the floor.
    The gray band represents the kinematically forbidden region.
    The bottom panel lies higher overall because the signal rate falls with $\delta$.
    }
    \label{fig:floor}
\end{figure}

Lowering the neutrino normalization uncertainties is commonly understood to lower the neutrino floor~\cite{Billard:2013qya,OHare:2021utq}.
This statement holds for the fog-derived floor for elastic scattering, but we show in Fig.~\ref{fig:floor} that it breaks down for inelastic dark matter.
When the uncertainty on the atmospheric rate is reduced from its nominal 25\% to 10\%, the fog-derived floor rises for every mass that can admit a $\delta=120$~keV splitting.
A 10\% uncertainty is comparable to the 7--9\% achieved by recent accelerator-tuned flux calculations below 1~GeV~\cite{Sato:2026avb}, and potentially testable with future low-energy atmospheric neutrino measurements~\cite{JUNO:2021tll,Newstead:2020fie}.

\begin{figure*}
    \centering
    \includegraphics[width=\textwidth]{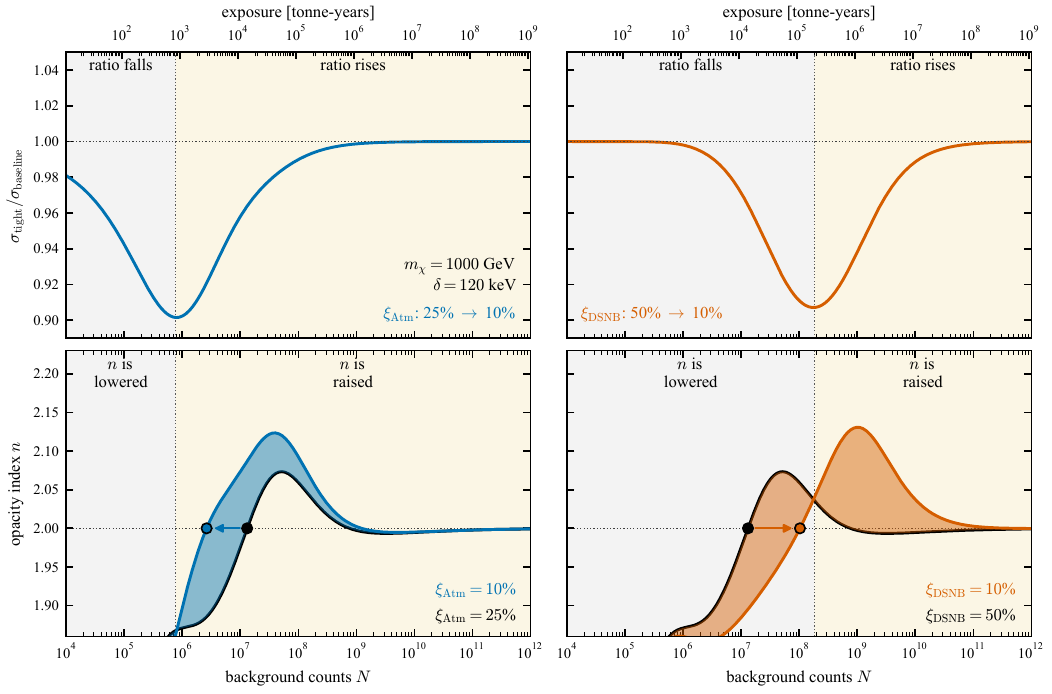}
    \caption{How reducing one background's normalisation uncertainty changes the opacity, at $m_\chi = 1000$\,GeV and $\delta = 120$\,keV (SHM).
    The left column shows the effect of reducing $\xi_{\rm Atm}$ from 25\% to 10\%, and the right column shows the effect of reducing $\xi_{\rm DSNB}$ from 50\% to 10\%.
    Top: ratio $r$ of the discovery limit with the reduced uncertainty to the baseline one, against total background counts $N$ (bottom axis) and exposure (top axis).
    The two limits agree at small $N$, where statistics dominate, and at large $N$, where the data constrain the normalizations, so $r$ falls to a minimum and rises back to unity.
    Bottom: opacity index $n$ with the baseline (black) and tightened (colour) uncertainty.
    Due to Eq.~\eqref{eq:opacity_ratio}, $n$ falls where $r$ falls (grey), and rises where $r$ rises (yellow).
    Dots mark the fog onset $N_\star$, at the first crossing of $n = 2$.
    Reducing $\xi_{\rm Atm}$ moves the onset to smaller exposures whereas reducing $\xi_{\rm DSNB}$ moves it to larger exposures.}
    \label{fig:connection}
\end{figure*}

The mechanism is as follows.
Reducing a background's rate uncertainty improves the discovery limit at all exposures.
However, the impact of improving on the uncertainty is minimal at small exposures where the statistical uncertainty dominates, and also at large exposures where high-count observations naturally provide a better constraint.
So the ratio of discovery limits $r=\sigma_\mathrm{tight}/\sigma_\mathrm{baseline}$ achieves a minimum at some intermediate exposure as shown in top row of Fig.~\ref{fig:connection}.
Using the definition of the opacity index (Eq.~\eqref{eq:opacity}),
\begin{equation}\label{eq:opacity_ratio}
\begin{split}
    -\frac{d \ln r}{d \ln N} = {} & - \frac{d \ln \sigma_\mathrm{tight}}{d \ln N} - \left( - \frac{d \ln \sigma_\mathrm{baseline}}{d \ln N}\right) \\
    = {} & \frac{1}{n_\mathrm{tight}} - \frac{1}{n_\mathrm{baseline}},
\end{split}
\end{equation}
which results in a decreased opacity at exposures below the location of the minimum in $r$, and an increased opacity at exposures above, as the normalization uncertainty tightens.
As shown in the bottom row of Fig.~\ref{fig:connection}, if the minimum in $r$ occurs before the onset of the fog as in the atmospheric tightening case, the floor rises, and vice-versa as in the DSNB tightening case.
Specifically for $\delta=120$~keV, the fog-onset is due to the DSNB opacity peak, and tightening the atmospheric uncertainty raises the opacity for exposures roughly above the location of the atmospheric peak, which itself remains below $n=2$.
This raises the floor.
The discovery limit at any fixed exposure still improves; the floor rises because the onset moves to a smaller exposure where the limit is higher.
Tightening the DSNB uncertainty, on the other hand, raises the opacity for exposures only well above the DSNB peak, and lowers the opacity at the location of the $n=2$ crossing, lowering the floor.

\section{Discussion}
\label{sec:discussion}

The results here are drawn from an idealized setup with unit efficiency down to $10^{-4}$~keV and perfect energy resolution, with neutrinos as the only backgrounds.
The threshold matters little for the onset of the fog for $m_\chi \gtrsim 100$~GeV, where changing the threshold to 3~keV reproduces the exposure needed to reach the fog to within a few percent.
We assume a static halo velocity distribution without the Earth's orbital motion, but this only has a few percent effect on the kinematic and fog edges.
We also take the atmospheric normalization to be constant in time, although the solar cycle modulates it by up to ${\sim}30\%$~\cite{Zhuang:2021rsg}.
At large $\delta$ the signal comes entirely from the high-speed tail of the velocity distribution.
Adding the LMC component moves the fog edge by only a few keV at 1~TeV (from 138 to 142~keV) but by about 30~keV near 100~GeV (from 144 to 176~keV), so the edge at low masses is the more sensitive to the halo model, which carries significant uncertainties~\cite{Green:2017odb}.

Both effects on the fog-derived floors discussed in Sec.~\ref{sec:floors} stem from how shape degeneracies interact with the opacity produced by multiple background sources with uncertain normalizations.
We suggest that future projections employing the opacity framework to calculate discovery limit floors verify that their limits are robust against both effects.
Because each source's uncertainty shapes its own feature in the opacity curve $n(N)$, the curve carries more robust information about the floor and allows these effects to be understood.
Discovery limit floors evaluated at fixed exposures remain smooth with shape parameters and do not rise when normalization uncertainties are lowered.

\section{Conclusion}
\label{sec:conclusion}

We have applied the opacity-based definition of the neutrino fog to inelastic dark matter scattering on xenon, scanning the dark matter mass $m_\chi$ and the mass splitting $\delta$ for the SHM and for a halo with additional high-velocity component associated with the LMC.
The mass splitting moves the signal to higher recoil energies, where the shape of the signal differs increasingly from that of the backgrounds.

The fog disappears well before the signal reaches the kinematic limit for dark matter heavier than around $m_\chi \gtrsim 100$~GeV.
For elastic scattering the fog arrives at about 70~tonne-years for $m_\chi \gtrsim 100$~GeV.
At $m_\chi=1$~TeV the opacity never exceeds $n=2$ for splittings beyond about 140~keV at any exposure up to $10^{10}$~tonne-years, while scattering remains kinematically allowed up to 395~keV (586~keV with the LMC component).
The mass splittings of a few hundred keV invoked to explain the LZ candidate lie entirely in the fog-free region.
For masses below around 50~GeV, in contrast, the fog persists almost to the kinematic limit.

Inelastic scattering reveals structure of the fog that is hidden in the elastic case.
The opacity as a function of exposure develops two peaks, set by the normalization uncertainties of the atmospheric neutrinos and the DSNB.
The onset of the fog is determined by the peak that first crosses the $n=2$ threshold.
This has two consequences for neutrino floors derived from the fog.
First, the floor can change discontinuously with any parameter that changes the shape of the signal.
Second, reducing the uncertainty on a single neutrino source can raise the floor: reducing the atmospheric uncertainty from 25\% to 10\% lowers the floor for elastic scattering but raises it by up to a factor 2 at $\delta = 120$~keV.
In both cases the discovery limit at any fixed exposure behaves as expected.

These results show that the neutrino fog is not an obstacle for inelastic dark matter searches with xenon at large splittings, and that floors derived from the fog opacity should be checked against fixed-exposure discovery limits, whenever more than one background is present alongside a shape-varying signal.

\begin{acknowledgments}
This work was funded by the US Department of Energy Grant No. DE-SC0011702. 
We thank N. Ezroura and H. Lippincott for helpful discussion.
\end{acknowledgments}

\makeatletter\immediate\write\@auxout{\string\citation{apsrev42Control}}\makeatother
\bibliographystyle{apsrev4-2}
\bibliography{refs}

\end{document}